\documentclass[conference]{IEEEtran}
\IEEEoverridecommandlockouts
\usepackage{cite}
\usepackage{amsmath,amssymb,amsfonts}
\usepackage{algorithm}
\usepackage{hyperref}
\usepackage{algorithmic}
\usepackage{graphicx}
\usepackage{textcomp}
\usepackage{subcaption}
\usepackage{amsmath}
\usepackage{empheq}
\usepackage{cases}
\usepackage{xcolor}
\def\BibTeX{{\rm B\kern-.05em{\sc i\kern-.025em b}\kern-.08em
    T\kern-.1667em\lower.7ex\hbox{E}\kern-.125emX}}
\begin{document}

\title{GSC: Generalizable Service Coordination\\
}

\author{\IEEEauthorblockN{Farzad Mohammadi}
\IEEEauthorblockA{\textit{School of ECE} \\
\textit{University of Tehran}\\
Tehran, Iran \\
mohammadi.farzad@ut.ac.ir}
\and
\IEEEauthorblockN{Vahid Shah-Mansouri}
\IEEEauthorblockA{\textit{School of ECE} \\
\textit{University of Tehran}\\
Tehran, Iran \\
vmansouri@ut.ac.ir}
}

\maketitle

\thispagestyle{plain}
\pagestyle{plain}

\begin{abstract}
	
Services with distributed and interdependent components are becoming a popular option for harnessing dispersed resources available on cloud and edge networks. However, effective deployment and management of these services, namely service coordination, is a challenging task. Service coordination comprises the placement and scalability of components and scheduling incoming traffic requesting for services between deployed instances. Due to the online nature of the problem and the success of Deep Reinforcement Learning (DRL) methods, previous works considered DRL agents for solving service coordination problems, yet these solutions have to be retrained for every unseen scenario. Other works have tried to tackle this shortcoming by incorporating Graph Neural Networks (GNN) into their solutions, but they often focus on specific aspects (and disregard others) or cannot operate in dynamic and practical situations where there is no labeled dataset and feedback from the network might be delayed. In response to these challenges, we present GSC, a generalizable service coordinator that jointly considers service placement, scaling, and traffic scheduling. GSC can operate in unseen situations without significant performance degradation and outperforms existing state-of-the-art solutions by 40\%, as determined by simulating real-world network situations. 

\end{abstract}

\section{Introduction}

Nowadays, with the increasing complexity of software systems and the proliferation of demand for resources, more services are deployed in a distributed fashion. The components of these services are often interdependent, meaning that the output of some units is used as the input of others. The flow of data inside these services can often be modeled using a Directed Acyclic Graph (DAG), and they are widely used in the industry. For example, microservices in a service mesh \cite{li2019service}, network slices in telecom networks \cite{afolabi2018network}, and machine learning pipelines \cite{verbraeken2020survey} are all distributed services with inter-dependent components.

On the other hand, to increase the accessibility of services and enhance their QoS, networks are growing at a fast rate and are decreasing their distance from end-user devices, leading to the edge computing paradigm \cite{ren2019survey}. As a result, networks have become extremely heterogeneous in terms of computing capacity, link bandwidth, and latency. Moreover, demands for services with diverse QoS requirements can enter networks from many locations that might be geographically distant. All of these can complicate the orchestration of these services to improve the performance metrics of the network, including throughput, utilization, and latency.

In order to correctly orchestrate these distributed services, several operations are performed. First, instances of service components, which can be virtual machines or containers, are placed at appropriate data centers in the network. Second, it is decided how many instances are required at each data center. Finally, incoming traffic is directed to the right function instance based on dynamic load. These three subproblems are regarded as placement, scaling, and scheduling, respectively. We should note that traffic scheduling is performed every time a component of a service computes its output since further computation might be continued at another data center for various reasons, such as QoS violations or lack of computing capacity at the current node.

Due to the online nature of the presented problem, Deep Reinforcement Learning (DRL) methods are a natural choice to solve it for multiple reasons. First, the intractability of optimization-based solutions for dynamic and time-sensitive problems, like our network orchestration problem. Second, modeling subproblems jointly is almost impossible without simplifying assumptions, which leads to solutions not suited for real networks. This also rules out the applicability of heuristic solutions to this problem. Third, DRL methods can handle a large number of states due to the generalizability of Deep Neural Networks (DNN). Fourth, DRL methods can upgrade themselves through online interaction with the environment, and as a result, there is no need for large labeled datasets. Finally, DRL-based solutions have low inference latencies, which are required in real-world networking solutions.

Despite the adoption of DRL in the literature for service coordination problems \cite{8931775, schneider2020selfdriving, schneider2021self}, to the best of our knowledge, all of them either do not consider all aspects of service coordination (placement, scaling, scheduling), or they lack generalizability to unseen network topologies and conditions. These works have to retrain their agents for every topology, and most of them for every set of node capacities and traffic patterns. Generalizability is essential for any practical solution to this problem since online training in deployment environments is time-consuming and network conditions change rapidly, so the coordinator must perform reasonably until the next round of training. A practical use case for this kind of service coordinator is AI-enabled Kubernetes \cite{k8s} controllers of distributed services. These controllers can be deployed in any cluster with any network topology and for many reasons retraining them in the deployment setup might be unfeasible.

In recent years, Graph Neural Networks (GNN) \cite{scarselli2008graph} have been applied to many problems in the networking domain \cite{rusek2019unveiling, hope2021gddr, 9310275, almasan2022deep, beurer2022learning}, mainly because of their ability to utilize the strong inductive bias of the input graph, producing generalizable solutions. However, these solutions disregard some aspects of service coordination or assume certain conditions, making them impractical \cite{heo2020graph, heo2020reinforcement, kim2020graph, siyu2021energy, 9201405, hara2022deep}. To address these issues, we present GSC, a generalizable service coordinator to orchestrate services with inter-dependent components in a multi-cloud setup. Overall, our contributions are:

\begin{itemize}
	\item We develop a GNN embedder based on Neural Algorithmic Reasoning (NAR), enabling the coordinator to receive input data in graph format and infer useful features from it by exploiting the inductive bias present in the input data.
	
	\item Develop a DRL-based agent to jointly consider traffic scheduling, scaling, and placement of chained functions in dynamic environments without any prior knowledge regarding incoming traffic patterns and traffic ingress nodes. This agent does not depend on any heuristics that tend to degrade the performance of the agent and limit its generalizability abilities.
	
	\item We extensively evaluate GSC in different degrees of generalizability. Results show that GSC can outperform existing state-off-the-art by up to 40\% in terms of successful flow rate.
	
	\item Last but not least, we make the source code of GSC publicly available to advocate reproducibility \cite{gsc}.
\end{itemize}

The structure of the paper is as follows: Sec. \ref{sec:related} introduces several works related to ours. Sec. \ref{sec:description} defines the details of service coordination and the problem setup. Sec. \ref{sec:gsc} presents GSC in detail. Sec. \ref{sec:eval} evaluates GSC and Sec. \ref{sec:conclusion} concludes the paper.

\section{Related works}
\label{sec:related}

\subsection{Generalization Efforts in Network Management and Orchestration}
One of the first works that investigated the idea of generalizing to unseen scenarios was RouteNet \cite{rusek2019unveiling}. In this work, the authors created a model for the prediction of QoS metrics, such as end-to-end delay and jitter, based on a given traffic matrix and routing policy in networks using GNN. They showed their model is able to operate on unseen network topologies effectively. In \cite{hope2021gddr}, authors developed a data-driven DRL-based agent for reducing network congestion. This GNN-enabled agent iteratively produced actions to accommodate different network topologies with unseen number of nodes and edges. In \cite{9310275}, a digital-twin model is created using GNN to predict the end-to-end delay of network slices, assisting the network orchestrator in preventing Service Level Agreement (SLA) violation. Authors of \cite{almasan2022deep} used GNN with DRL to effectively route lightpaths in optical transport networks, showcasing the idea of in-observation actions to allow their agent to generalize to different network topologies. Beurer-Kellner et al. \cite{beurer2022learning} used Neural Algorithmic Reasoning (NAR) \cite{velivckovic2021neural} to train a model for the protocol-agnostic configuration of computer networks. Their solution also works on much larger networks.

\subsection{GNN-enabled Service Coordination}
Heo et al. \cite{heo2020graph} developed an encoder-decoder model using GNN to place SFCs in a network to minimize delay and maximize the success rate of deployment. However, their supervised approach limited the applicability of their solution to real-world scenarios and they also did not consider scalability. Authors extended this work in \cite{heo2020reinforcement} and adopted DRL to solve the labeling challenge, but both works do not consider traffic scheduling among service components. Kim et al. \cite{kim2020graph} developed a classifier using GNN to select an optimal number of VNF instances, yet their solution is not applicable to dynamic setup and does not consider traffic redirection. Siyu et al. \cite{siyu2021energy} minimize energy consumption and delay using a GNN-enabled DDQN agent. However, they do not consider traffic flows and operating on other networks than the one the agent is trained on. DeepOpt \cite{9201405}  uses a GNN-assisted DRL agent in the SDN controller to effectively place VNFs in the network along with considering generalizability, yet they don't consider any form of traffic engineering while solving their problem. Contrary to the mentioned works, Hera et al. \cite{hara2022deep} focus on finding service paths for SFCs using GNN, dismissing SFC placement and scaling.

The most relevant work to GSC is DeepCoord \cite{schneider2020selfdriving, schneider2021self}, which also inspires us. DeepCoord considers joint placement, traffic scheduling, and placement using periodic monitoring data and does all of these using a DDPG-based agent (without GNN). The biggest disadvantage of DeepCoord is the lack of generalizability abilities to unseen network topologies and scenarios. GSC solves the mentioned problem in DeepCoord by incorporating GNN and introducing an efficient DRL agent design.

\section{Problem Description}
\label{sec:description}

\subsection{Multi-cloud Network Formulation}

We model the multi-cloud network as an undirected graph $G=(V, E)$ comprising a set of vertices $V$ and a set of edges $E$. Each vertex represents a data center in the network, which from now on we call a node, and every edge models the communication link between two data centers, such as an MPLS link or a lightpath from the underlying optical network. Each node, or data center, has a finite set of computing capabilities $C_v \in R^{n_v}$ that represents the available CPU, GPU, etc. $n_v$ is the number of computing capabilities associated with each node and for the sake of simplicity, in this work, we only consider one computing capability, meaning $n_v =1$, but more than one capability can be used. Moreover, each link has an available bandwidth $BW_l$ and a delay $D_l$, which are affected by the distance between the source and destination nodes.

Services are modeled as a chain of functions $s_i=\{f_{i, 1}, f_{i, 2}, ..., f_{i, n}\}$, each of which can be a container, a virtual function, or even a serverless function. Demands for services are modeled as incoming flows. Each incoming flow $F_i = (s_i, src_i, r_i, t_i)$ arrives at an ingress node $src_i$ at time $t_i$ requesting service $s_i$ with the data rate $r_i$. To be successful, $F_i$ should be processed by the functions of the requested service $\{f_{i, 1}, f_{1, 2}, ..., f_{i, n}\}$ in order.  $f_{i, n}$ is the $n$'th function inside the chain defining service $s_i$. If a flow cannot be scheduled to an appropriate destination node or the destination cannot process the flow, then the flow will be dropped.

In this work, we consider delayed feedback and monitoring data, from the network similar to \cite{schneider2021self}. This kind of observation is aligned with real-world network monitoring systems that are used in the deployment, such as Prometheus Stack \cite{prometheus}, a renowned open-source monitoring solution. In this way, we only observe \textit{cumulative} monitoring data periodically. For example, the CPU usage of data centers is not known at any given time step. Instead, we observe cumulative CPU usage in the last monitoring period $MP$, after the end of $MP$. In addition to the status of nodes, such as resource usage, we receive cumulative link and ingress traffic observations from the network.

\subsection{Decision Variables and Objective}
\label{sec:variables}

We define two decision variables to model the service coordination problem. The first decision variable is the \textit{scheduling tensor} $x \in R^{n \times |V| \times |V|}$ where $n$ is the maximum length of function chains and $|V|$ is the number of nodes in the network. Each element $x_{i,j,k} \in [0, 1]$ is the probability of scheduling flow $F_\tau$ that needs to be processed by function $f_{\tau, i}$, which belongs to service $s_\tau$, from node $V_j$ to node $V_k$. From the above definition, we should have:

\begin{equation}
	\label{eq1}
	\sum_{k=1}^{|V|}{x_{i,j,k}} = 1 \quad  \forall i \in \{1, 2, ..., n\},\; \forall j \in \{1, 2, ..., |V|\}
\end{equation}

Note that this formulation can easily be extended to support multiple chained services at once by adding another dimension to the scheduling tensor, yet for the sake of simplicity, we only consider one service.
The second decision variable is a matrix called \textit{deployment indicator} $y \in R^{\mathcal{F} \times |V|}$ where $\mathcal{F}$ is the maximum number of functions that need to be deployed in the network. Each element $y_{i, j} \in \{0, 1\}$ indicates whether function $f_i$ is deployed at $V_j$ or not.

Until now, we modeled scheduling and placement problems by two decision variables. Regarding the third problem, scaling, we turn our focus to inter-node scalability since most modern data centers exploit resource orchestration software that automatically handles intra-node scalability, such as Kubernetes \cite{k8s} and OpenStack \cite{openstack}. Moreover, we note that the inter-scalability problem can be modeled by the deployment indicator by allowing multiple columns of a single row, which represents a single function, to be set to 1. This means a single function is deployed at multiple data centers, thus achieving inter-node scalability.

Now we state that the \textit{deployment indicators} can be determined using a scheduling tensor and a \textit{deployment policy}. This policy maps the scheduling tensor to the deployment indicator. Again, for the sake of simplicity, we use a simple policy that sets $y_{i, j}$ to 1 whenever $x_{i, \tau, j}, \forall \tau \in \{1, 2, ..., |V|\}$ is greater than 0. 

In this work, we consider the rate of successful flows scheduled over the network as our objective, which is defined in (\ref{eq:obj}):

\begin{equation}
	\label{eq:obj}
	Obj = \frac{\psi_{succ} - \psi_{drop}}{\psi_{succ} + \psi_{drop}} \in [-1, 1]
\end{equation}

In the above equation, $\psi_{succ}$ and $\psi_{drop}$ are the number of successful and dropped flows in the last $MP$, respectively.

\section{Generalizable Service Coordination}
\label{sec:gsc}

Here we present design aspects of GSC. First, in Sec. \ref{sec:overview} we describe the overall method and challenges, then we start discussing GNN Embedder at Sec. \ref{sec:embedder} and DRL agent at Sec. \ref{sec:agent}. Finally, we introduce the RL environment at Sec. \ref{sec:env}, and in Sec. \ref{sec:algorithm}, we present the algorithm used for training GSC.

\subsection{Overview}
\label{sec:overview}

Based on our discussion about decision variables and \textit{deployment policy}, we only need to calculate scheduling tensor $x$ at every $MP$ to conduct service coordination. However, several aspects need to be considered before computing a scheduling tensor.

To choose a suitable action, DNN-based methods need to somehow embed the input information, which in our case is in the form of a graph, into a feature representation. Previous works \cite{schneider2020selfdriving, schneider2021self, saha2023deep}, embed features of nodes and links into a vector. This tends to pose two challenges. First, this way of embedding is not permutation-invariant. Therefore, even a single network can be embedded in many ways, complicating the learning process. Second, This naive representation dismisses the strong inductive bias that exists in graphs as interconnection. Using this bias is imperative for designing high-performance solutions.

The second aspect is that generalizable agents should be able to process input data with variable sizes since different networks have different numbers of nodes and edges. This challenge is hard to overcome since traditional DNNs, such as MLP and CNNs, require fixed input sizes. The third issue is that the size of the scheduling tensor, the ultimate output of the agent, is dependent on the number of nodes in the input network. Thus, the service coordinator should be able to produce outputs with variable sizes, which again cannot be accommodated using traditional models. In the next subsection, we will address the first and second challenges by introducing a graph embedder.

\subsection{GNN Embedder}
\label{sec:embedder}

\begin{figure}[t]
	\centering
	\includegraphics[width=0.45\textwidth]{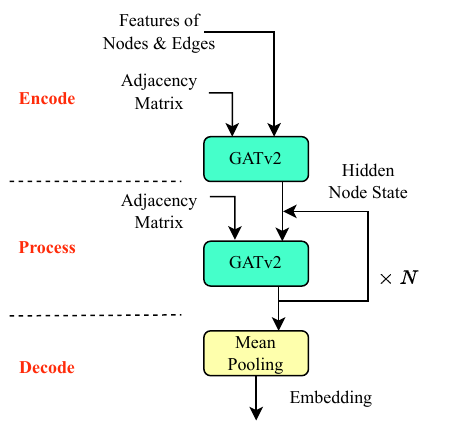}
	\caption{GNN Embedder}
	\label{fig:embedder}
\end{figure}

GNNs are a natural solution for generating an embedding for several reasons: they do not rely on ordering to produce embedding, making them permutation-invariant by design \cite{battaglia2018relational}, they strongly utilize the inductive bias existing in the input data \cite{battaglia2018relational}, leading to more accurate results, and they have shown the ability to extrapolate well for nonlinear tasks \cite{xu2020neural}. However, the generalizability of GNNs depends on the appropriate design of the architecture and components. This is challenging due to the remarkable expansion of GNN's designs space \cite{you2020design}, leading to the various choices for architecture \cite{gilmer2017neural, battaglia2018relational}, message passing layer \cite{li2015gated, kipf2016semi, velivckovic2017graph}, and pooling methods \cite{liu2022graph}.

Message Passing Neural Networks (MPNNs) \cite{gilmer2017neural} have shown a great potential in previous works \cite{rusek2019unveiling, hope2021gddr, pujol2021ignnition, hara2022deep}. Thus, we select MPNN as our choice for the architecture of GNN Embedder. MPNN can be expressed by the following equations:

\begin{eqnarray}
	\label{eq:message}
	&\textrm{Message}: \quad m_{vw}^{t+1} = M(h_v^t,h_w^t,e_{v,w}) \\
	\label{eq:agg}
	&\textrm{Aggregation}: \quad m_v^{t+1} = A(\{ m_{vw}^{t+1} : w \in N(v)\}) \\
	\label{eq:update}
	&\textrm{Update}: \quad h_v^{t+1} = U(h_v^t,m_v^{t+1})
\end{eqnarray}

In (\ref{eq:message}), $h_i^t$ represents the hidden state of node $i$, which is initialized with the features of node $v_i$ at $t=0$. $e_{v,w}$ represent the features of the edge connecting $v_v$ and $v_w$. Function $M$ is the message creation that should be chosen. (\ref{eq:agg}) employs permutation-invariant aggregation function $A$, operating on incoming messages to node $v_v$ from its neighbors $N(v)$ and (\ref{eq:update}) generates the next representation for node $v_v$ using update function $U$. The choice of $M$, $A$, and $U$ defines a specific message-passing layer.

Among many proposed message-passing layers in previous works, Graph Attention Networks (GAT) \cite{velivckovic2017graph} are particularly more successful at generalizing to unseen scenarios, as shown in works like \cite{beurer2022learning}, mainly due to their attention mechanism \cite{knyazev2019understanding}. However, authors of \cite{brody2021attentive} show that the GAT layer uses static attention that can hamper the learning process. Instead, they propose a new layer, GATv2, which solves the GAT problem using a simple modification. In our design, we use GATv2 as the GNN layer. In (\ref{eq:gat_message}), (\ref{eq:gat_agg_1}), (\ref{eq:gat_agg_2}), and (\ref{eq:gat_update}), we can see how GATv2 specifies \textit{Message}, \textit{Aggregation}, and \textit{Update} operations:

\begin{equation}
	\label{eq:gat_message}
	\textrm{Message}: \; m_{vw}^{t+1} = a^T LeakyReLU(W.[h_v^t || h_w^t || e_{v,w}])
\end{equation}

In (\ref{eq:gat_message}), $a$ and $W$ are learnable parameters and $||$ is the concatenation operation.

\begin{numcases}{\textrm{Aggregation}: \quad}
	\alpha^{t+1}_{v, w} = softmax_w(m_{vw}^{t+1})&  \label{eq:gat_agg_1}\\
	m_v^{t+1} = \sum_{w \in N(v)}{\alpha^{t+1}_{v, w}Wh_w^t}& \label{eq:gat_agg_2}
\end{numcases}

Note that in (\ref{eq:gat_agg_1}) and (\ref{eq:gat_agg_2}), $\alpha$ is the attention variable, which acts as the weighting coefficient.

\begin{equation}
	\label{eq:gat_update}
	\textrm{Update}: \quad h_v^{t+1} =  \alpha_{v,v}Wh_v^t + m_v^{t+1}
\end{equation}

Regarding the architecture of GNN Embedder, inspired by recent advances in NAR \cite{velivckovic2021neural} to tackle various combinatorial problems \cite{cappart2023combinatorial}, we adopted the \textit{encode-process-decode} architecture to enable our embedder to reach convergence by iterative message passing steps in the \textit{process} part of the architecture.

The complete design of the GNN Embedder can be seen in \figurename \ref{fig:embedder}. We chose a simple pooling layer, mean pooling since we are operating on relatively small networks. This Embedder produces embeddings that are permutation-invariant, generalizable, and fixed-sized regardless of the size of the input graph. Hence, effectively solving the first two issues discussed in Sec. \ref{sec:overview}. The final issue, variable action space, will be addressed in the next section.

\subsection{DRL Agent}
\label{sec:agent}

\begin{figure}[t]
	\centering
	\includegraphics[width=0.45\textwidth]{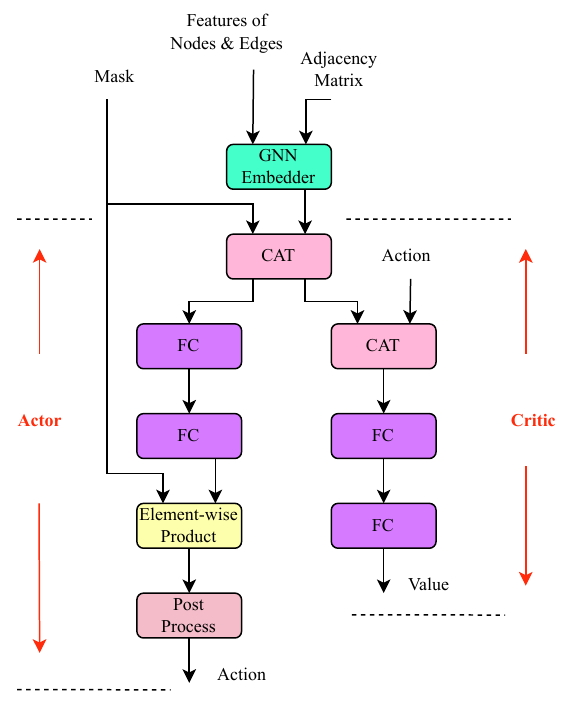}
	\caption{DRL agent}
	\label{fig:agent}
\end{figure}

Being generalizable means generating \textit{scheduling tensors} with variable sizes corresponding to the input network topology. On the other hand, DRL methods mainly employ DNN models that have fixed output sizes. To circumvent this challenge a possible solution is to \textit{iteratively} generate parts of the output, used by previous works \cite{addanki2019placeto, hope2021gddr}. The main obstacle in incorporating this idea is the difficulty of enforcing (\ref{eq1}). Particularly, every value in a row of scheduling tensor is dependent on other values in that row. One naive idea might be to enforce (\ref{eq1}) after every complete episode, where the entire scheduling tensor has been generated, but when the input topology changes, the DRL agent cannot keep track of the logic behind this manually added operation.

A better solution is \textit{masking} the action space during generation of actions \cite{mirhoseini2020chip}. In this way, we can fix the size of the action space in a way that it can accommodate the largest possible network topology and mask invalid (or unused) parts of the generated output. The biggest advantage of this \textit{masking} solution over \textit{iterative} method is that (\ref{eq1}) can be easily enforced and we can help the DRL agent to learn the masking pattern by providing the mask itself for the DRL agent. Therefore, we have chosen this idea for the design of the DRL agent. To mask a generated output, we generate a boolean mask tensor with the same size as \textit{scheduling tensor}, and then take the element-wise product of the generated output by the Actor network and mask tensor.

Since both action and observation space are continuous in our problem, we use the DDPG algorithm \cite{lillicrap2015continuous}, in which the DRL agent is composed of Actor and Critic networks. Actor network chooses a suitable action receiving required inputs, and the Critic network trains the Actor network based on the received feedback from the monitoring system. \figurename \ref{fig:agent} depict the complete DRL agent. Here, FCs are Fully connected layers and CAT operations perform concatenation. As we can see, the mask is provided to both Actor and Critic networks, enabling them to learn the pattern of masking. Moreover, we apply the mask by using element-wise product operation between the mask and the Actor's output, treating the mask like a hard constraint. Finally, we enforce (\ref{eq1}) in the Post Process block.

\subsection{RL Environment}
\label{sec:env}

Here, we introduce the Partially Observable Markov Decision Process (POMDP) environment in which GSC operates.

\subsubsection{Observation Space}
The observation space of the RL environment is comprised of four components: nodes' features, edges' features, adjacency matrix, and mask. Features of nodes include $C_v$, cumulative ingress traffic, and cumulative ratio of used resources. For edges, we have two features: $D_l$ and cumulative used bandwidth. All of these features are normalized to [-1, 1]. Adjacency matrix has the shape of $2\times |E|$, and the mask is expressed in a tensor with the size of $n \times |V|_{max} \times |V|_{max}$.

\subsubsection{Action Space}
The action space of the environment is the maximum \textit{scheduling tensor} possible. Therefore, we have chosen $|V|_{max} = 64$. Hence, action space is a tensor with the shape of $n \times 64 \times 64$.

\subsubsection{Reward}
In section \ref{sec:variables}, we defined our objective in (\ref{eq:obj}). We directly take that objective as our reward signal, which is calculated periodically.

\subsection{Training Algorithm}
\label{sec:algorithm}

\begin{algorithm}[t]
	\caption{Training algorithm}
	\label{alg1}
	\begin{algorithmic}[1]
		\STATE agent $\gets$ GSC()
		\STATE env $\gets$ ENV()
		\STATE obs $\gets$ env.reset()
		\FOR {$i = 1$ to $L \times N_{EP}$}
		
		\STATE action $\gets$ agent.choose\_action(obs)
		\STATE action $\gets$ post\_processing(action)
		\STATE nx\_obs, reward, done = env.step(action)
		\STATE store\_in\_buffer(obs, nx\_obs, reward, done)
		\STATE obs = nx\_obs
		\IF{$TP$ condition is true}
		\IF{i $\ge W$}
		
		\FOR{$\eta$ times}
		\STATE batch $\gets$ sample\_from\_buffer()
		\STATE agent.train\_critic(batch)
		\STATE agent.train\_actor(batch)
		\STATE agent.update\_target\_networks()
		\ENDFOR
		
		\ENDIF
		
		\ENDIF
		
		\ENDFOR 
	\end{algorithmic} 
\end{algorithm}

\begin{figure*}[h]
	\centering
	\begin{subfigure}[t]{0.23\textwidth}
		\centering
		\includegraphics[width=\textwidth]{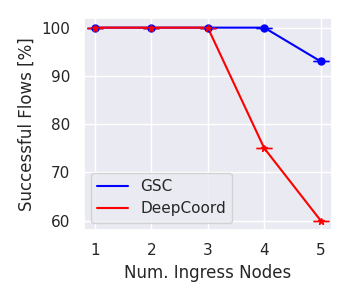}
		\caption{Fixed Arrival}
		\label{fig:perf-1}
	\end{subfigure}
	\hfill
	\begin{subfigure}[t]{0.23\textwidth}
		\centering
		\includegraphics[width=\textwidth]{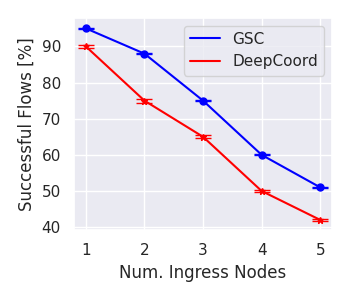}
		\caption{Poisson Arrival}
		\label{fig:perf-2}
	\end{subfigure}
	\hfill
	\begin{subfigure}[t]{0.23\textwidth}
		\centering
		\includegraphics[width=\textwidth]{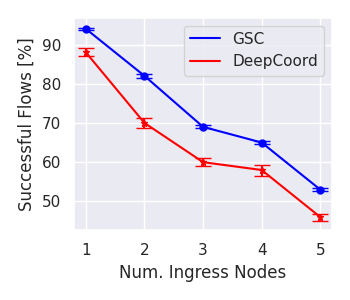}
		\caption{MMPP Arival}
		\label{fig:perf-3}
	\end{subfigure}
	\hfill
	\begin{subfigure}[t]{0.23\textwidth}
		\centering
		\includegraphics[width=\textwidth]{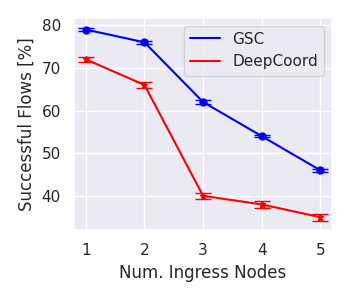}
		\caption{Real-world Traffic}
		\label{fig:perf-4}
	\end{subfigure}
	\caption{Performance Evaluation with \textit{\textbf{seen}} scenarios.}
	\label{fig:seen_scenraios}
\end{figure*}

Algorithm \ref{alg1} is used for training GSC. Before any training, we initialize the agent and environment (ln. 1-3). The number of iterations in this algorithm is determined by the number of episodes $N_{EP}$ and the length of each episode $L$ (ln. 4). At each iteration, agent produces an action by adding Gaussian noise to the output value of actor network (ln. 5). The noise is essential for agent's ability to learn since it promotes exploration. To ensure the condition described by (\ref{eq1}), we apply the post-processing function (ln. 6). In this post-processing, we ensure that the entries of scheduling tensor are greater than a \textit{scheduling threshold} to prevent sending small fractions of traffic to destination nodes. Then, we calculate new reward and observation through simulation (ln. 7). To enable faster and more effective learning, we store transitions in a graph experience buffer (ln. 8). Once the condition for training period $TP$ is met (for example every episode) and number of passed iteration is greater than a warm-up period $W$ we start training agent (ln. 10-11). We will repeat training for $\eta$ gradient steps (ln. 12). In each gradient step, we get a mini-batch from graph experience buffer (ln. 13) and start training all four neural networks of the agent according to \cite{lillicrap2015continuous} (ln. 14-16). Note that the scheduler automatically resets the environment every $L$ time step and changes its parameters according to the scheduling configuration, so there is no need for manually resetting it.

Since graphs can have an arbitrary number of vertices and relations, traditional ways of mini-batching samples are infeasible or not memory efficient. Therefore, we need to use another approach of mini-batching in graph learning. As suggested in \cite{fey2019fast}, we stack the adjacency of different samples in a diagonal fashion in a larger matrix and concatenate features of vertices.

\section{Implementation}
In this part, we discuss the implementation details of GSC and the surrounding platform required for its training and inference.

\subsection{Simulator and Environments}
\label{sec:sim}

We based our simulator on CoordSim \cite{coordsim}, which itself is based on Python and SimPy \cite{meurer2017sympy}. Using this simulator, we create a Gym \cite{towers_gymnasium_2023} environment to streamline the training process. The central part of this environment is a scheduler that periodically changes the configuration of the network topology, including features of entities and permutation of ingress nodes. This scheduler itself can be configured via a YAML file. For the observation, we use the standard data structure proposed by  PyG \cite{fey2019fast}, which accommodates the adjacency matrix, mask, and features of entities.

\subsection{Algorithm and Agent}
\label{sec:alg}

To enable easier manipulation, we based our DRL agent on CleanRL \cite{huang2022cleanrl}. CleanRL provides a research-friendly implementation of DRL algorithms, which is based on PyTorch \cite{Paszke_PyTorch_An_Imperative_2019}. Suggested replay buffers used in this implementation cannot store graph data structures with variable sizes. As a result, we have developed a graph replay buffer issue.

\section{Evaluation}
\label{sec:eval}

Here, we evaluate GSC agents from various facets. Section \ref{sec:eval_setup} details the setup for holistic performance measurement, Sec. \ref{sec:seen_scenario} outlines experiments conducted in seen scenarios, and Sec. \ref{sec:unseen_scenario} assesses the generalizability of GSC in unseen situations.

\subsection{Evaluation Setup}
\label{sec:eval_setup}

We rely on real-world network topologies, acquired from the internet topology zoo \cite{knight2011internet}, to conduct our experiments. These topologies merely determine the nodes and their connectivity, so we needed to add several features, such as node capacity and link bandwidth, to them. To enable reproducibility, instead of determining these parameters randomly at run time, we chose a set of random parameters once for each topology and updated it with those values. Note that we have generated multiple sets of these parameters to mimic various scenarios, including bandwidth-constrained networks. To be specific, we chose four networks from Internet Zoo to conduct our experiments: Claranet, Compuserve, BtEurope, and Abilene. The first three networks are used for training GSC and the last topology, Abilene, is used for inference.

Since prior works in \cite{schneider2020selfdriving, schneider2021self} show the significant advantage of DeepCoord over other solutions, such as \cite{draxler2018scaling}, we choose DeepCoord as our main baseline. In this work, the authors considered a service chain with three network functions. Therefore, we perform our own experiments in a similar setup. For determining inter-arrival times between flows, we consider four scenarios: Fixed, Poisson, Markov-modulated Poisson Process (MMPP) \cite{fischer1993markov} real-world traffic scenarios from SNDlib \cite{orlowski2010sndlib}.

To infer our trained agent, throughout all of our experiments, we test the agent over 20000 time steps and set the monitoring period $MP$ to 100. The scheduling threshold is 0.1  and we repeat every experiment 25 times to calculate the amount of error indicated by error bars in depicted figures.

During all experiments, node features are ingress traffic, node load, and node capacity. Features of edges are delay and used bandwidth. The length of each episode is 200. The buffer limit is set to 10000. The mean and standard deviation of random noise used for exploration in the algorithm are 0.15 and 0.3, respectively. Both actor and critic use 200 steps to warm up without training. we set $\tau$ to 0.0001 for updating actor and critic softly. Finally, $\gamma$ and the learning rate are set to 0.99 and 0.001, respectively.

Inside the GNN Embedder, we use a single GATv2 layer for the encoder and another GATv2 layer for the processor. The hidden dimension is 64 throughout the GNN Embedder and the number of iterations at the processor is 4. For an actor, we use a single FC layer with 256 nodes and an FC layer with 64 nodes for the critic. All activation layers are ReLU. Policy update frequency for the DDPG algorithm is (1, "episode"), meaning that we update both actor and critic after every episode. In this process, we use mini-batches of size 100.

\begin{figure*}[h]
	\centering
	\begin{subfigure}[t]{0.46\textwidth}
		\centering
		\includegraphics[width=\textwidth]{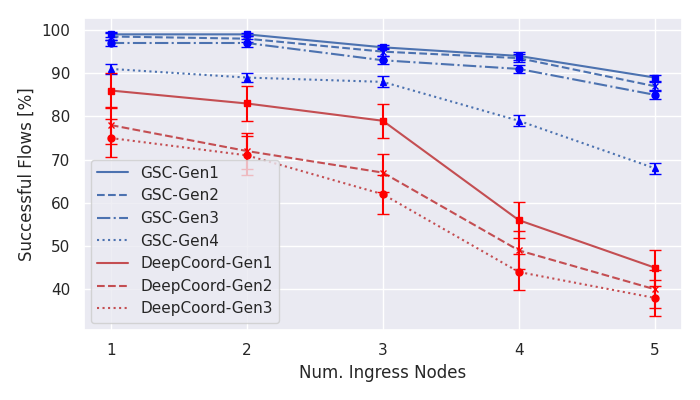}
		\caption{Fixed Arrival}
		\label{fig:gen_1}
	\end{subfigure}
	\hfill
	\begin{subfigure}[t]{0.46\textwidth}
		\centering
		\includegraphics[width=\textwidth]{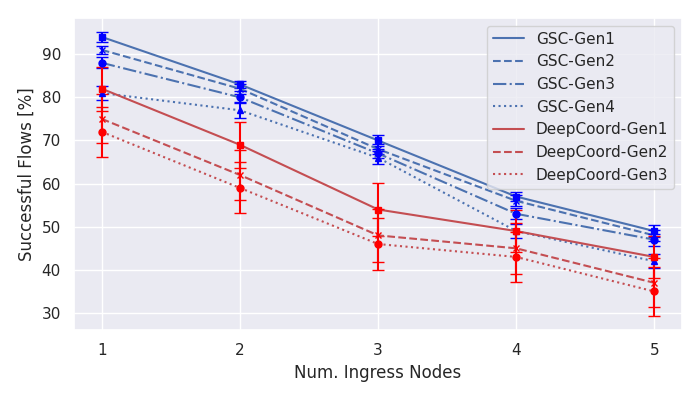}
		\caption{Poisson Arrival}
		\label{fig:gen_2}
	\end{subfigure}
	\hfill
	\begin{subfigure}[t]{0.46\textwidth}
		\centering
		\includegraphics[width=\textwidth]{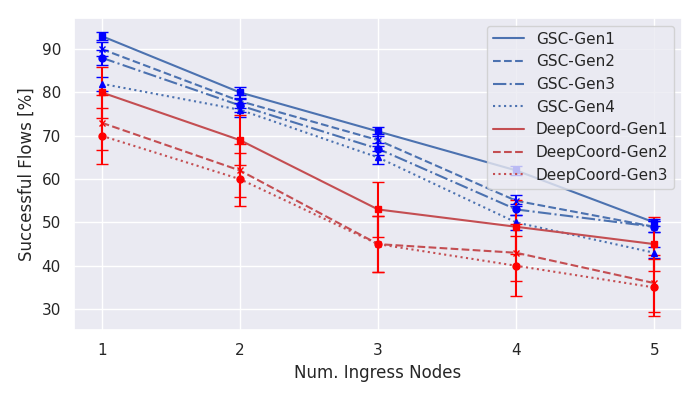}
		\caption{MMPP Arival}
		\label{fig:gen_3}
	\end{subfigure}
	\hfill
	\begin{subfigure}[t]{0.46\textwidth}
		\centering
		\includegraphics[width=\textwidth]{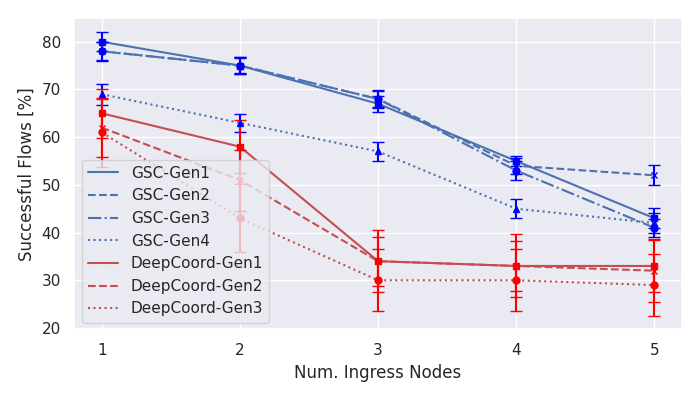}
		\caption{Real-world Traffic}
		\label{fig:gen_4}
	\end{subfigure}
	\caption{Performance Evaluation with \textit{\textbf{unseen}} scenarios}
	\label{fig:unseen_scenarios}
\end{figure*}

\subsection{Performance Evaluation over Seen Scenarios}
\label{sec:seen_scenario}

First, we perform a simple one-to-one comparison between GSC and DeepCoord to capture the significance of GNN Embedder in the performance of GSC. In these experiments, whose results are depicted in \figurename \ref{fig:seen_scenraios}, we train both GSC and DeepCoord on a specific scenario, in terms of the number and permutation of ingress nodes, the capacity of nodes, and network topology. After reaching convergence, we evaluate them on the exact scenario and report the results. We have considered four traffic generation models and for each model, we progressively increase the number of ingress nodes to gauge the ability of agents in high-demand situations.

\figurename \ref{fig:perf-1} shows for fixed inter-arrival cases, both solutions achieve the same results, but after 4 ingress nodes, GSC performs significantly better, indicating its ability to operate in resource-constraint scenarios more efficiently. This is mainly due to the fact that GNN Embedder is able to factor in neighbors of nodes during message passing iterations, enabling nodes to decide better actions for resource-constrained scenarios imposed by excessive demand. Note that errors in both solutions are negligible since this scenario lacks randomness.

\figurename \ref{fig:perf-2} depicts the ability of GSC to outperform DeepCoord consistently in every number of ingress nodes. The dynamic attention mechanism of GATv2 layers enables nodes to dynamically focus on different sets of neighbors at run-time, which is extremely useful in these scenarios because inter-arrival times are generated from Poisson distributions. Apart from minor performance degradation and increased error compared to Poisson arrival, both solutions in \figurename \ref{fig:perf-3} perform similarly to Poisson setup.

From \figurename \ref{fig:perf-4}, we can deduce that the performance gap between these two solutions increases as we add ingress nodes in real-world traffic scenarios. This shows the ability of GSC to capture more intricate and dynamic setups through the use of NAR to reason with better precision.

\subsection{Performance Evaluation over Unseen Scenarios}
\label{sec:unseen_scenario}

We devise four types of generalization and we evaluate the performance of GSC at every combination traffic pattern, number of ingress nodes, and generalization type. We consider four generalization dimensions:

\begin{enumerate}
	\item Capacity of nodes
	\item Permutation of ingress nodes
	\item Number of ingress nodes
	\item Network topology
\end{enumerate}

We sequentially add these dimensions to define generalization types to see how solutions perform in various steps, in which each step is more challenging than the previous, toward full generalization. These types are

\begin{enumerate}
	\item Type 1 (Gen1): dimension 1
	\item Type 2 (Gen2): dimension 1 \& 2
	\item Type 3 (Gen3): dimension 1 \& 2 \& 3
	\item Type 3 (Gen4): dimension 1 \& 2 \& 3 \& 4
\end{enumerate}

To make these types more clear, consider the following example. Suppose we want to test the agent in Gen2, which is generalization type 2. This agent sees the exact network topology and number of ingress nodes used for inference during the training process, but it doesn't see the evaluation network capacity and the permutation of ingress nodes considered for inference.

For training, based on the exact generalization type, agents see various values for defined generalization dimensions except for inference values. We change these values periodically to allow agents enough time to adapt to new scenarios.

Since the action space of DeepCoord is fixed, we are not able to test Gen4 using this agent. This is because the change of network topology directly translates to a change of action space size in most cases (due to the differences in the number of nodes). In \figurename \ref{fig:unseen_scenarios}, we see the result of testing both agents in various generalization types.

According to \figurename \ref{fig:gen_1}, the performance of GSC is not much affected through Gen1 to Gen3, but after adding the network topology dimension, Gen4, the performance decreases noticeably, which is expected because the topology dimension is extremely challenging for GSC to master. On the other hand, the major drop for DeepCoord happens between Gen1 and Gen2 since FC layers cannot capture different permutations of nodes. Moreover, the error is significantly lower with GSC, thereby indicating the stability of this solution. Furthermore, GSC even in the most difficult scenario, Gen4, outperforms DeepCoord in the simplest type, Gen1, showing the effectiveness of the design.

As expected, \figurename \ref{fig:gen_2} and \figurename \ref{fig:gen_3} show similar results. Compared to the fixed arrival scenario, we can see that GSC manages to keep the error low, about the same as before, yet DeepCoord's standard deviation increases. Again, this indicates the abilities of GNNs to operate in stochastic problems. Finally, \figurename \ref{fig:gen_4} displays the remarkable advantage of GSC over DeepCoord in real-world traffic scenarios due to its robustness and precision.

\subsection{Other aspects}
Despite the numerous benefits that GSC provides in service coordination, it has several downsides too. For instance, training GSC takes longer because more neural network layers are used in it, which also complicates the hyperparameter tuning process. This can be mitigated by using methods like meta-learning \cite{finn2017model}. Furthermore, with the advent of new computing paradigms like Fog computing \cite{tuli2023ai}, centralized training procedures will not be effective. Distributed learning approaches must be adopted to make solutions like GSC applicable to these paradigms.

\section{Conclusion}
\label{sec:conclusion}

In this work, we present GSC, a GNN-enabled DRL solution, for coordinating chained services in multi-cloud networks. We designed a GNN Embedder to enable our DRL agent to operate in unseen network topologies without significant performance degradation. Our extensive evaluations show that GSC offers DRL-based solutions both in seen and unseen scenarios, thus paving the way for AI-enabled network orchestration.

For future works, we propose three directions: Using more advanced GNN architectures to produce more efficient embedders, utilizing methods like meta-learning to enable GSC to self-adapt in online setups, and adopting distributed training algorithms. 

\bibliographystyle{IEEEtran}
\bibliography{ref}

\end{document}